\documentclass[fleqn,12pt,twoside]{article}
\usepackage{espcrc1,epsfig}

\usepackage{graphicx}
\usepackage[figuresright]{rotating}

\newcommand{\AmS}{{\protect\the\textfont2
  A\kern-.1667em\lower.5ex\hbox{M}\kern-.125emS}}

\title{The Lattice Fermi Surface}

\author{Simon Hands\address{Department of Physics, University of Wales
Swansea,\\
Singleton Park, Swansea SA2 8PP, U.K.}
        \thanks{Supported by EU contract ERBFMRX-CT97-0122 and by the
Leverhulme Trust.}}
\begin{document}

\maketitle

\begin{abstract}
The Nambu -- Jona-Lasinio model in 2+1 dimensions is simulated for non-zero
baryon chemical potential with a diquark source term. 
No evidence for a BCS condensate or gap is seen at high density; rather, 
critical behaviour with novel exponents is observed, suggesting that 
$2d$ superfluidity as first described by Kosterlitz and Thouless is realised,
but with the universality class determined by the presence of relativistic 
fermions.
\end{abstract}

\section{INTRODUCTION}

\begin{figure}[htb]
\begin{center}
\epsfig{file=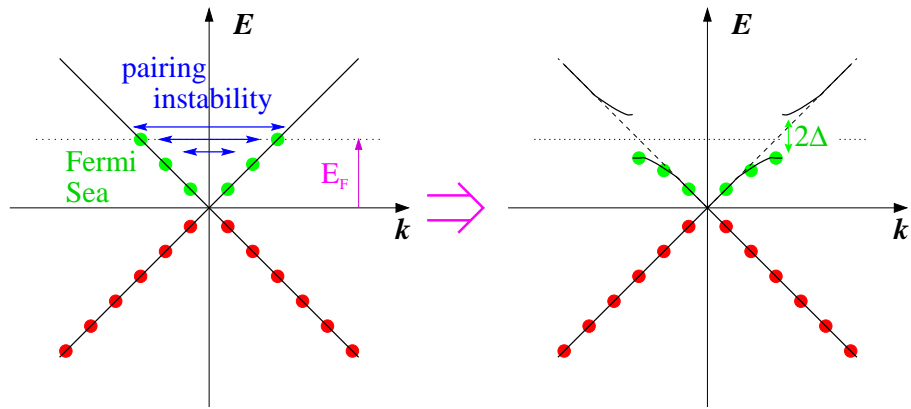, width =12cm}
\end{center}
\caption{
\label{fig:disperse}
The BCS instability in degenerate matter.}
\end{figure}
There has been intense recent speculation on the form of the QCD phase diagram 
at small temperature $T$ but large baryon chemical potential $\mu$.
The conventional view is that for
sufficiently large $\mu$, chiral symmetry is spontaneously
restored, resulting in a degenerate system of relativistic quarks known as 
quark matter, with Fermi energy $E_F=\mu/3$. Since the $qq$ interaction can be
attractive due to gluon exchange \cite{BBL} and/or instanton effects
\cite{RSSV}, however,
this simple picture
may have a BCS instability with respect to 
condensation of diquark pairs at the Fermi
surface, resulting in a color superconducting ground state \cite{ARW},
and the formation of an energy gap $\Delta\not=0$ 
to the first excited
state of the system (see Fig.~\ref{fig:disperse}).
Model calculations \cite{BR}
suggest that $\Delta$ could be as large as 100 MeV, comparable with the
constituent quark scale.

It is clearly desirable to examine the various scenarios
in non-perturbative lattice QCD simulations, where 
systematic control is at least in principle possible. Unfortunately, this has so
far proved impracticable because the Euclidean functional measure is no longer
positive definite for $\mu\not=0$, rendering the usual
importance sampling methods
ineffective. There are, however, simpler four-fermion models 
where Monte Carlo methods can be applied and yield a plausible description of
degenerate matter \cite{Hands}. In this talk I describe numerical studies
of the Nambu -- Jona-Lasinio (NJL) model in 2+1 spacetime dimensions; the
Lagrangian in continuum notation reads
\begin{equation}
{\cal L}=\bar\psi({\partial\!\!\!/\,}+\mu\gamma_0+m)\psi-{g^2\over2}
\left[(\bar\psi\psi)^2-(\bar\psi\gamma_5\vec\tau\psi)^2\right].
\end{equation}
The model in 3+1$d$
has a long history as an effective description of strong interaction
physics \cite{NJL}. For coupling $g^2$ stronger than some critical $g_c^2$
the SU(2)$_L\otimes$SU(2)$_R\otimes$U(1)$_B$ global symmetry present for 
quark mass $m\to0$ spontaneously breaks to SU(2)$_I\otimes$U(1)$_B$,
accompanied by the generation of a constituent quark mass
$\Sigma=g^2\langle\bar\psi\psi\rangle$. The behaviour in
2+1$d$ has the same pattern, except that this time there is an interacting 
continuum limit at $g^2\to g_c^2$, $\Sigma/\Lambda\to0$ \cite{Hands}.
Once a chemical potential is introduced, for low $T$ an expansion in the inverse
number of flavors $1/N_f$ predicts $\Sigma(\mu)$ to remain 
roughly unchanged out to a critical $\mu_c\simeq\Sigma(0)$, whereupon 
chiral symmetry is abruptly restored in a strong first-order transition
\cite{KRWP}. At the same point the baryon density
$n_B=\langle\bar\psi\gamma_0\psi\rangle$ rises from zero to a non-zero
$n_B\propto\mu^2\theta(\mu-\mu_c)$, 
consistent with filling a two-dimensional Fermi ball of radius $\mu$. 
This is confirmed by
lattice simulations \cite{Hands}; in particular, unlike other simulations
with a real measure, there is a clear separation of scales between
$\mu_c$ and $m_\pi$ \cite{pi}.

\section{THE DIQUARK SECTOR}

The question to consider is whether for $\mu>\mu_c$ the baryon number
U(1)$_B$ symmetry is spontaneously broken by a diquark condensate
$\langle qq\rangle\not=0$. Since the NJL model is not a gauge theory, this
leads not to superconductivity but to the associated phenomenon of
{\sl superfluidity\/}. In fact, numerical studies \cite{HM}
reveal that the preferred
channel for pairing in this regime is (in the notation of staggered fermions)
via the scalar SU(2)$_L\otimes$SU(2)$_R$ singlet $\chi^{tr}\tau_2\chi$.
To find unambiguous evidence for BCS condensation on a finite system, however,
the correct procedure \cite{MH} is to add to ${\cal L}$ a diquark source term
\begin{equation}
j_\pm qq_\pm\equiv j_\pm(\chi^{tr}\tau_2\chi\pm\bar\chi\tau_2\bar\chi^{tr}).
\label{eq:source}
\end{equation}
It is then possible
to measure the diquark condensate using methods similar to those used in 
lattice studies of chiral symmetry breaking:
\begin{equation}
\langle qq_+\rangle={1\over V}{{\partial\ln Z}\over{\partial j_+}}
\end{equation}
together with the associated susceptibilities
\begin{equation}
\chi_\pm=\sum_x\langle qq_\pm(0)qq_\pm(x)\rangle.
\end{equation}
In a conventional BCS condensation the + describes a ``Higgs'' mode while the --
is a ``Goldstone'', which is constrained by a Ward identity:
\begin{equation}
\chi_-(j_-\!\!=0)={{\langle qq_+\rangle}\over j_+}.
\label{eq:ward}
\end{equation}

\begin{figure}[htb]
\begin{minipage}[t]{80mm}
\epsfig{file=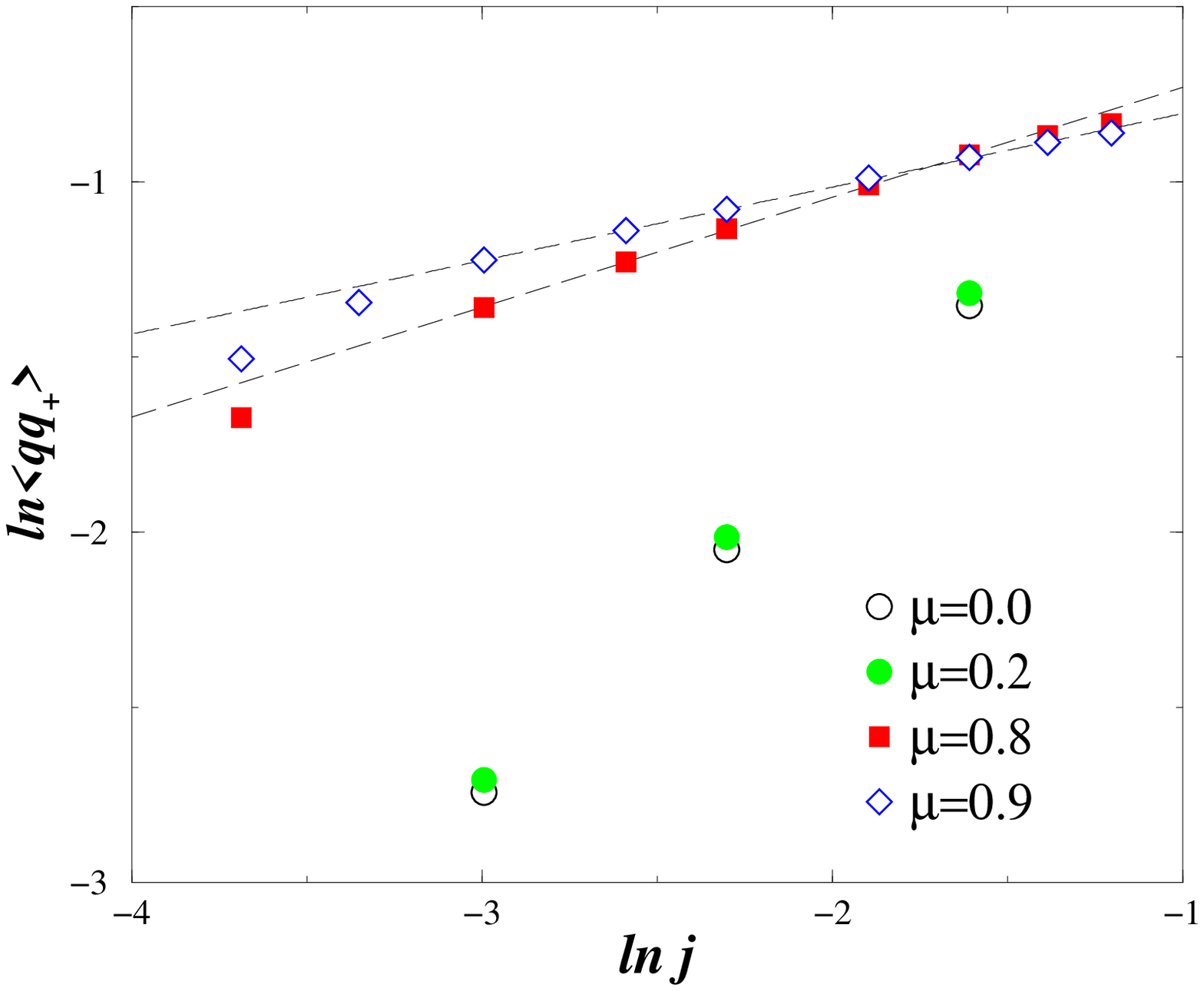, width =8cm}
\caption{$\ln\langle qq_+\rangle$ vs. $\ln j$.}
\label{fig:lnln}
\end{minipage}
\hspace{\fill}
\begin{minipage}[t]{80mm}
\epsfig{file=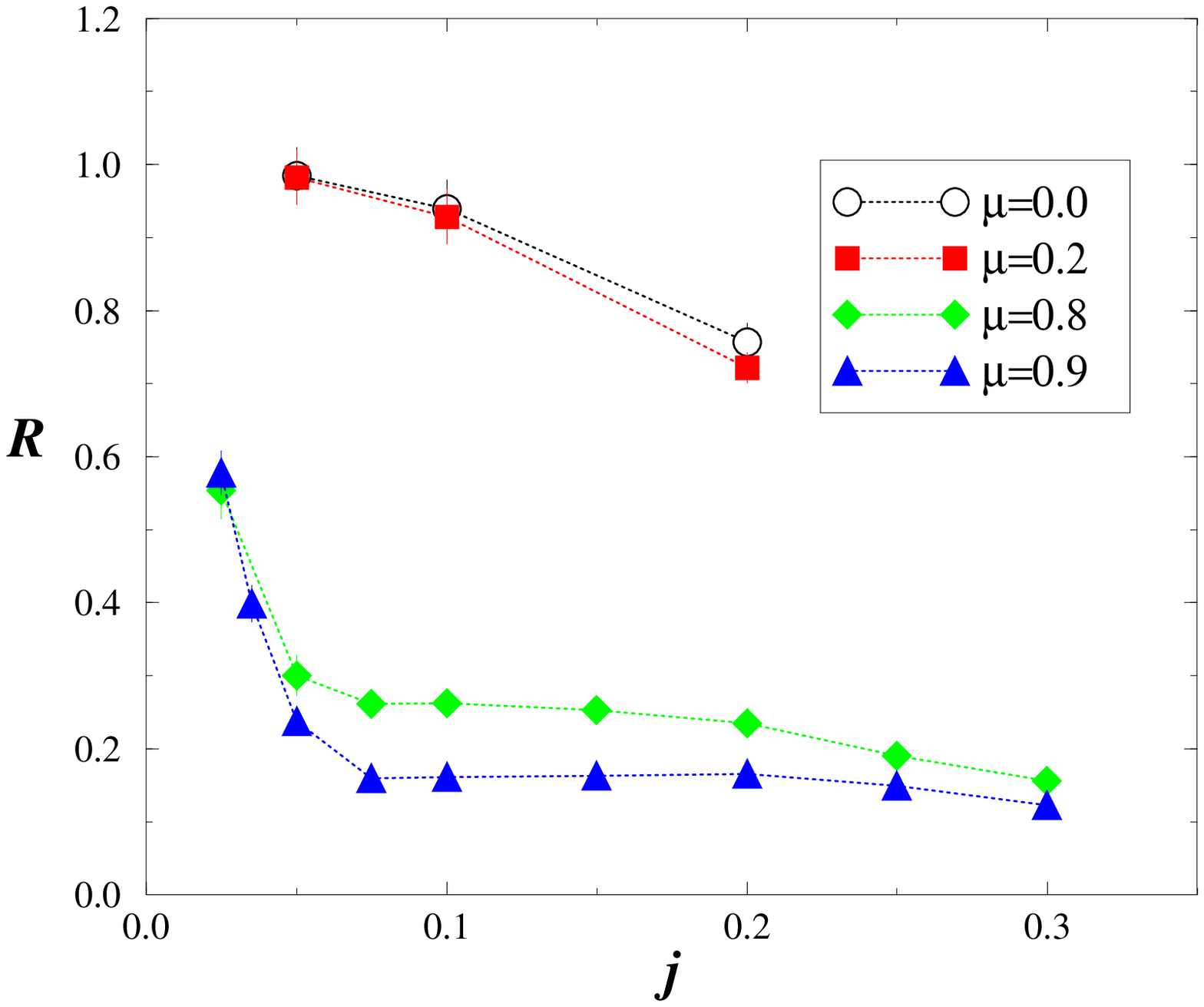, width =8cm}
\caption{$\vert\chi_+/\chi_-\vert$ vs. $j$.}
\label{fig:R}
\end{minipage}
\end{figure}
In collaboration with Biagio Lucini and Susan Morrison I 
have simulated a lattice NJL model including the source term
(\ref{eq:source}) \cite{HLM}. Our parameter choice yields a 
$\mu_c\simeq0.65$.
Fig.~\ref{fig:lnln} plots $\langle qq_+\rangle$ extrapolated to the 
zero temperature $L_t\to\infty$ limit; for $\mu<\mu_c$ the data support a linear
relation between $\langle qq_+\rangle$ and $j$, consistent with a
U(1)$_B$-symmetric ground state. For $\mu>\mu_c$ by contrast, the data suggest
\begin{equation}
\langle qq_+\rangle\propto j^\alpha
\label{eq:power}
\end{equation}
with $\alpha=\alpha(\mu)$ falling in the range 0.2 - 0.3 for the $\mu$ values
studied. Evidence for power law behaviour is reinforced by considering the 
susceptibility ratio $\vert\chi_+/\chi_-\vert$: eqns. (\ref{eq:ward}) and
(\ref{eq:power}) together imply that this ratio should take the constant
value $\alpha$ in the high density phase, and the plateaux of Fig.~\ref{fig:R}
for $\mu=0.8,0.9$ support this.

The simulation data suggest that the high density phase $\mu>\mu_c$ is
{\sl critical\/}, characterised by continuously varying exponents $\delta(\mu)
=\alpha^{-1}$ and
$\eta(\mu)$ defined via
\begin{equation}
\langle qq(0)qq(\vec x)\rangle\propto{1\over{\vert\vec x\vert^\eta}},
\;\;\;\;\;\;\;\;\;\vec x\in\mbox{R}^2,
\label{eq:phase}
\end{equation}
and is thus qualitatively similar to the low-$T$ phase of 
the $2d$ XY model.
If we write $qq(x)=\phi_0e^{i\theta(x)}$, then
long range order is washed out by IR fluctuations of $\theta$,
but long-range phase coherence persists via (\ref{eq:phase}). This is precisely
what gives rise to persistent currents and hence superfluidity in $2d$ systems. 
The supercurrent $\vec J_s$
is related to $qq$ via
\begin{equation}
\vec J_s=K_s\vec\nabla\theta.
\end{equation}
The only way to change the circulation $\kappa=\oint\vec
J_s.\vec{dl}$
around a periodic volume is to create a vortex -- anti-vortex in the 
$\{\theta\}$ configuration  
and translate one of the pair around the universe in the
perpendicular direction until they reannihilate, in so doing incrementing
the quantised $\kappa$ by $2\pi K_s/L$. 
The energy needed increases logarithmically
with $L$, implying that the circulation is metastable \cite{KT}.

\section{THE SPIN-${1\over2}$ SECTOR}

\begin{figure}[htb]
\begin{minipage}[t]{80mm}
\epsfig{file=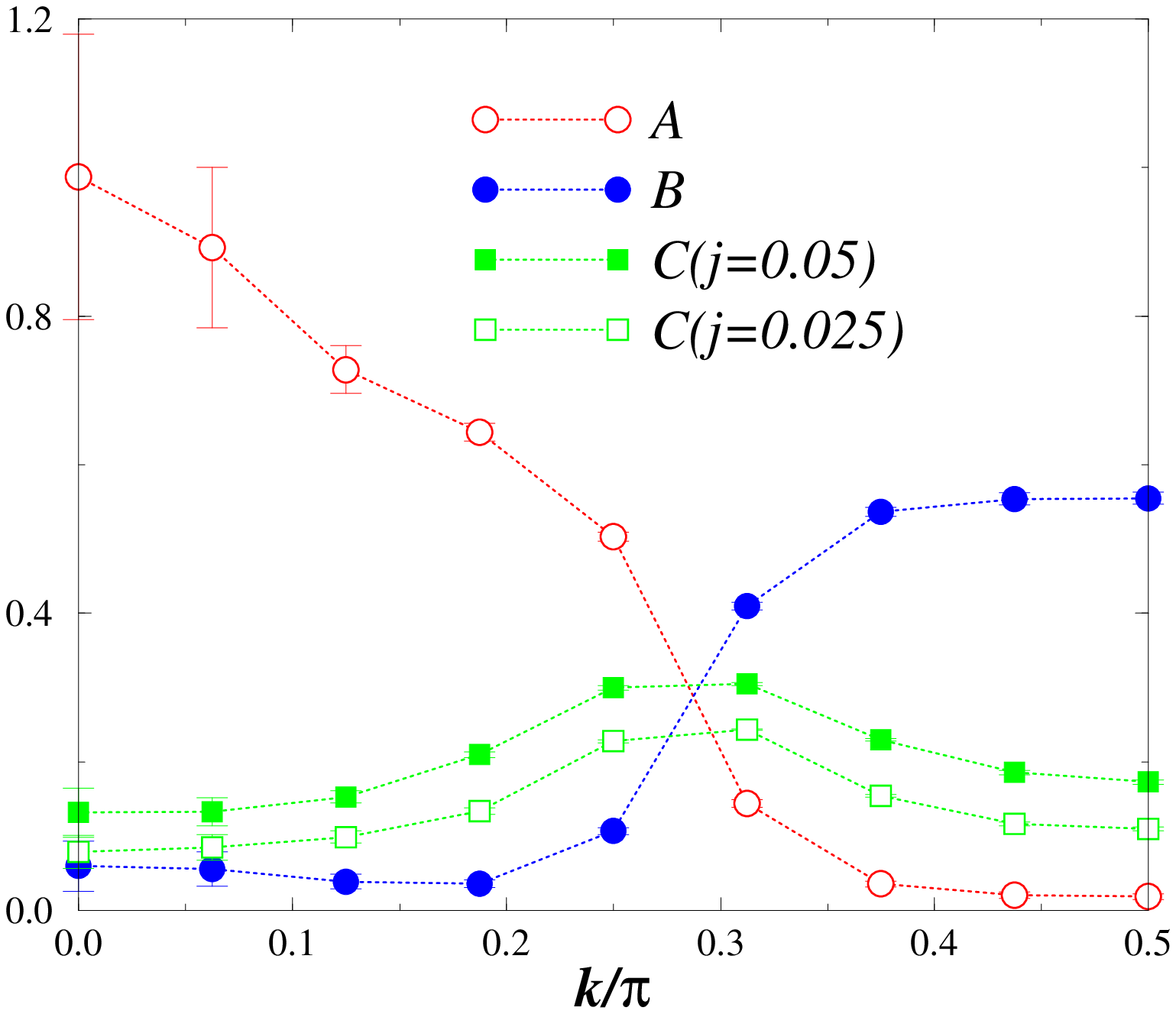, width =7cm}
\caption{Amplitudes for hole ($A$),\hfill\break
particle ($B$) and anomalous ($C$)
propagation.}
\label{fig:amplitudes}
\end{minipage}
\hspace{\fill}
\begin{minipage}[t]{80mm}
\epsfig{file=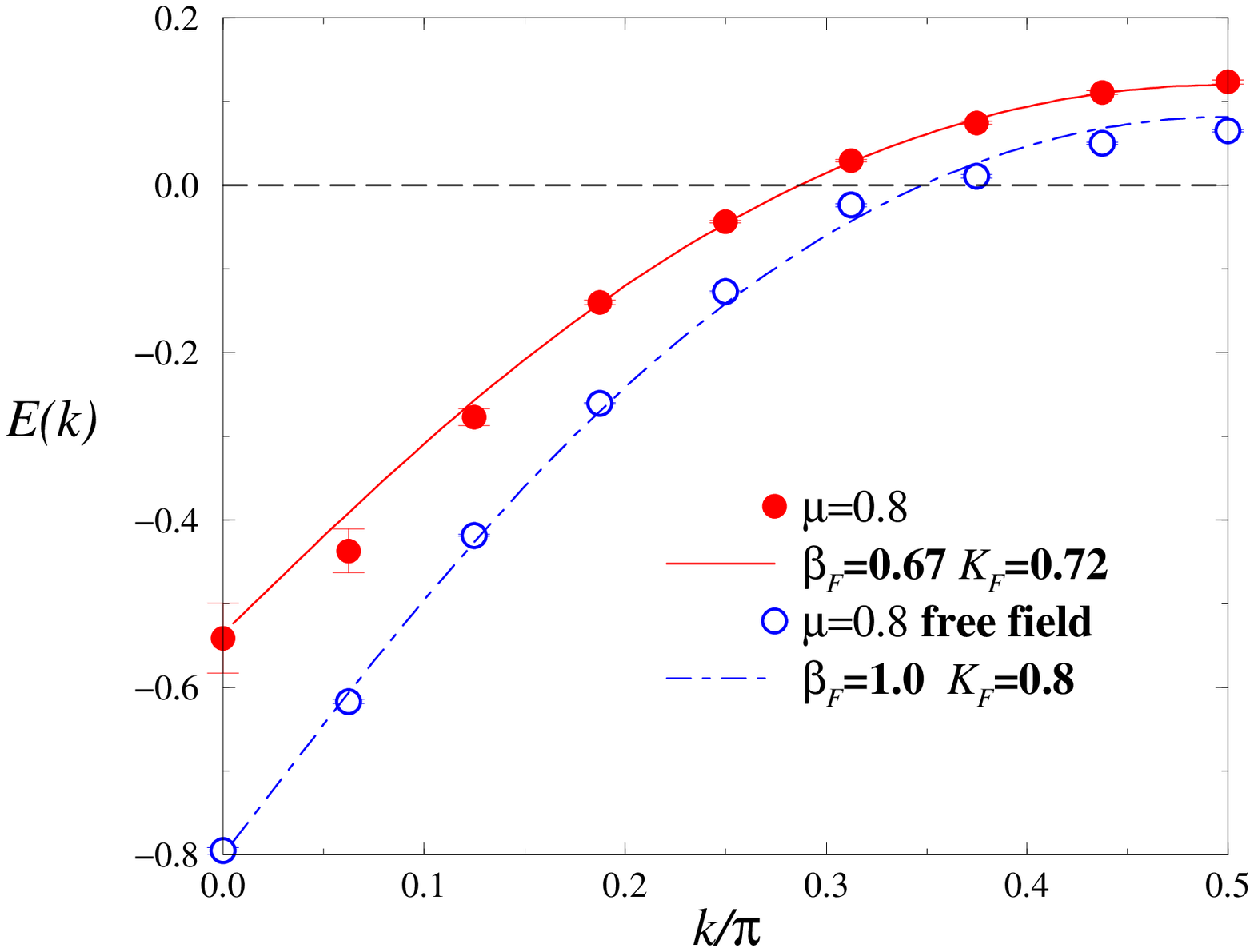, width =8cm}
\caption{Dispersion relation $E(k)$ for both free and interacting
fermions.}
\label{fig:eofk}
\end{minipage}
\end{figure}
The conclusions of the previous section beg the question of why the observed
$\delta\simeq3$ -- 5 is not consistent with the $2d$ XY value
$\delta(T)\geq15$, in contrast 
to a recent study of superfluidity in the attractive
Hubbard model \cite{CO}. 
Further insight is gained from studying the
spin-${1\over2}$ sector via the fermion propagator, which for $j\not=0$ contains
both normal $\langle q(0)\bar q(x)\rangle$ and anomalous $\langle
q(0)q(x)\rangle$ components. To probe the Fermi surface, data at non-zero
momentum $\vec k$ are needed. In the normal sector a sharp transition
between forwards $Ae^{-Et}$ and backwards $Be^{-E(L_t-t)}$ propagation at 
a rather well-defined momentum $k/\pi\simeq0.28$ on a $32^3$ system at
$\mu=0.8$, as shown in Fig.~\ref{fig:amplitudes}. We interpret this as
a crossover between hole and particle excitations at the Fermi surface.
The anomalous propagators signifying particle-hole mixing peak 
at the same momentum, but appear to vanish as $j\to0$.

The results for the mass gap are shown in Fig.~\ref{fig:eofk}, with hole
energies plotted as negative. The resulting curve is the {\sl quasiparticle
dispersion relation\/}; its detailed form indicates a Fermi momentum $k_F$
slightly less than $\mu$ and a Fermi velocity $\beta_F={\partial E}/{\partial
k}\vert_{k=k_F}\simeq0.7$, somewhat less than the speed of light. Both
results differ from the corresponding free-field values, but are
consistent with a relativistic normal Fermi liquid with repulsive interaction
between quasiparticles with parallel momenta \cite{BC}. Most importantly, there
is no signal for a gap $\Delta\not=0$. The origin of the non-standard critical
behaviour may therefore be attributed to massless fermions;
the presence of a Fermi surface leads to a new $2d$ universality class. This
situation should be contrasted with dimensional reduction in systems with $T>0$
\cite{PW}; 
because the fermi degrees of freedom in this case acquire
masses $\pi T$ and decouple, 
the universality class is that of the spin model with
the corresponding global symmetry.

\section{CONCLUSION}

Both $\langle qq\rangle$ and $\Delta$ vanish in the limit
$j\to0$ implying that the conventional BCS description appropriate
for, say,
superfluid $^3$He,  does not apply in this case. 
The most economic hypothesis explaining the results of Figs.~2 -- 5 is that 
the NJL$_{2+1}$ model in its high density phase describes a gapless relativistic
system with superfluidity realised 
in the way first suggested by Kosterlitz and Thouless for 
thin films of $^4$He. 
Further numerical confirmation for this picture would come from observation
of current quantisation in the presence of a source $j(x)$ with built-in winding
number. The more urgent problem, however, is to extend the 
calculations to NJL$_{3+1}$ and test the exciting scenarios predicted in
\cite{RSSV,ARW,BR} beyond the self-consistent approximation.

\end{document}